\pdfoutput=1

\documentclass{agujournal2018_DKoronczay}

\usepackage{url}

\usepackage{aas_macros}

\journalname{JGR-Space Physics}

\begin{document}

\renewcommand{\APACrefYearMonthDay}[3]{\APACrefYear{#1}} % disable months in the reference list by apacite

\title{VLF transmitters as tools for monitoring the plasmasphere}

\authors{
D\'avid Koronczay\affil{1,2},
J\'anos Lichtenberger\affil{1,2},
Lilla Juh\'asz\affil{1},
P\'eter Steinbach\affil{3,1}, 
George Hospodarsky\affil{4}
}

\affiliation{1}{Department of Geophysics and Space Sciences, Eotvos University, Budapest, Hungary}
\affiliation{2}{Research Centre for Astronomy and Earth Sciences, Hungarian Academy of Sciences, Sopron, Hungary}
\affiliation{3}{MTA-ELTE Research Group for Geology, Geophysics and Space Sciences, Budapest, Hungary}
\affiliation{4}{Department of Physics and Astronomy, University of Iowa, Iowa City, IA, USA}

\correspondingauthor{David Koronczay}{david.koronczay@ttk.elte.hu}

\begin{keypoints}
\item Satellite observation of ducted VLF transmitter signals in the plasmasphere, based on wave characteristics
\item Propagation inversion method of ducted VLF transmitter signals yield electron densities in agreement with reference measurements
\item Can be an alternative tool for measuring plasmaspheric electron densities
\end{keypoints}

	\begin{abstract}
 Continuous burst mode VLF measurements were recorded on the
 Van Allen Probes satellites and are analyzed to detect
 pulses from the Russian Alpha (RSDN-20) ground-based
 navigational system between January and March 2016.
 Based on the wave characteristics of
 these pulses and on the position of the spacecraft, the signals
 propagated mostly in ducted mode in the plasmasphere. Knowledge of the
 propagation path allowed us to carry out a monochromatic wave
 propagation inversion to obtain plasmaspheric electron
 densities. We compared the obtained densities with
 independent in-situ measurements on the spacecraft. The
 results show good agreement, validating our inversion
 process. This contributes to validating the field-aligned
 density profile model routinely used in the inversion of whistlers
 detected on the ground. Furthermore, our method can provide
 electron densities at regimes where no alternative
 measurements are available on the spacecraft.
 This raises the possibility of using
 this method as an additional tool to measure and monitor plasmaspheric
 electron densities.
\end{abstract}

%% ------------------------------------------------------------------------ %%
%
%  TEXT
%
%% ------------------------------------------------------------------------ %%

\section{Introduction}

Artificial very low frequency (VLF, 3 to 30 kHz) signals have long been a useful tool in plasmaspheric research. They have been shown by ground-based measurements to be able to propagate in ducts in the magnetosphere \citep{mcneill67}. They were first observed in space by low Earth orbit satellites Ariel-3 and Ariel-4 that recorded the signals of
various NATO naval transmitters (specifically GBR, NWC and NAA),
both above the transmitters and at their respective conjugate regions, suggesting ducted propagation \citep{bullough69,bullough1975}. Similar low Earth orbit conjugate observations were reported by \citet{cerisier73} of the FUB transmitter by the FR-1 satellite, by
\shortciteA{larkina83}
in an experiment observing a Soviet transmitter at L=2.6 by the Intercosmos-19 satellite and by \citet{sonwalkar94} in another experiment observing signals on the Cosmos-1809 satellite sent from a transmitter in Khabarovsk. On the other hand, observations of unducted VLF transmitter signals on FR-1 were reported by \citet{cerisier74}, and later on the high altitude satellites Explorer-45 and Imp-6 by
\shortciteA{inan77},
ISEE-1 by
\shortciteA{bell1981},
GEOS-1 by
\shortciteA{neubert83}
and DE-1 and Cosmos-1809 by
\citet{sonwalkar1986} and \citet{sonwalkar94}, originating
variously from Siple, Omega, Alpha and Khabarovsk transmitters. In these experiments, no six-component measurements of the electric and magnetic fields were available, and thus various assumptions were necessary to attempt to determine the wave characteristics. \citet{shawhan} lists further early examples of such multicomponent measurements.
%\citet{yamamoto91}
\shortciteA{yamamoto91}
provides a determination of wave normal and Poynting vectors of an Omega signal observed by the Akebono satellite, based on five-component measurements.
\textbf{}
Artificial VLF waves were proposed as a tool for studying wave-particle interactions, most famously in the Siple experiment \citep{sipleexperiment}. \citet{nunn2015} gives a short review of research into triggered emissions. It was also shown that VLF transmitter signals may influence electron precipitation \citep{imhof1983}. \citet{smith87} were able to infer the L-value of the path and the cross-drift velocity of ducts from ground-based Doppler measurements of the NAA and NSS transmitters.

\citet{inan77} attempted a reconstruction of the electron density distribution of the plasmasphere based on raytracing and the measured group delays of unducted VLF pulses in a satellite observation by the Imp~6 satellite. The lack of complete wave characteristics, and the reliance on theoretical models for the field-aligned density distribution make a robust reconstruction challenging. Even more importantly, the fact that unducted whistlers can follow difficult to predict, complicated and often multiple paths, with sometimes multiple solutions, makes a general solution
-- i. e. inversion of observed unducted signals --
elusive.

\shortciteA{kimura2001} proposed a similar procedure, using wave normal directions and propagation delay times of unducted pulses from Omega transmitters (operational until 1997) observed on the Akebono satellite, compared to raytracing of the signals and fitting the parameters of a diffusive equilibrium model of the plasmaspheric densities. The obtained values have not been compared with independent density measurements.

With a view to previous results of both ducted and unducted VLF transmitter pulses, \citet{clilverd2008} investigated the relative importance of the two types of propagation. They came to the conclusion that for low L-shell transmitters ($L<1.5$), significant portion of the wave energy propagating into the plasmasphere is nonducted, while for larger L-shells, waves become highly ducted. The interhemispherically ducted propagation is limited by the minimum electron half gyrofrequency along the propagation path \citep{smith_etal1960} (imposing an upper limit in L-value, depending on the frequency). \citet{song2009} conducted a targeted monitoring of ground based VLF transmitters by the IMAGE satellite between 2001 and 2005 which also supported the propagation of these signals along the magnetic field lines.

The ubiquitous existence of ducted VLF signals makes it possible to use the group delay of such signals to measure electron densities in the plasmasphere. For a ducted signal detected by a satellite, we can understand its propagation path. Thus, a relatively straightforward propagation inversion can be carried out to obtain the electron density. This is also facilitated by recently available, more precise, experimental models of the electron density profiles along the field lines \citep{ozhogin2012}.

The Van Allen Probes satellites have been in orbit since 2012, with 6-channel wave experiments onboard (Electric and Magnetic Field Instrument and Integrated Science, EMFISIS, \citet{kletzing2013}) that are appropriate for detecting and confirming ducted VLF signals. We are focusing on the Alpha VLF transmitters, which emit short pulses, with gaps between the pulses, making it relatively easy to determine the arrival time of each pulse. The relationship of their frequencies and the L-values of the transmitters permit the signals to travel the full interhemispheric path in most of the plasmasphere, thus, they should be detectable in a wide range of spacecraft positions, as discussed in Section \ref{section_transmitters}.

This paper describes the methodology of using observations of such VLF pulses for determining plasmaspheric electron densities.
The utility of such a procedure is twofold. Firstly, it validates the propagation inversion process, and more specifically, the electron density profile models. This same inversion process is used, for example, for whistler inversion, a standard tool for acquiring plasmaspheric
electron densities. It forms the basis of a recent global ground-based network for automatic detection and inversion of whistlers (AWDANet, see
\shortciteA{lichtenberger2008}
and \citet{lichtenberger2010}). Second, this method escapes some of the limitations of other electron density measurements,
e.g. those derived from plasma resonance frequencies measured by EMFISIS on the Van Allen Probes satellites \citep{kurth2015}
and by WHISPER on the CLUSTER satellites \citep{decreau2001,kougblenou2011}.
Thus, it may possibly be used to regularly extend such density measurements, without any additional investment.

In Section \ref{section_transmitters} we review the information
we gathered on the Alpha transmitters, while in Sections \ref{emfisis}-\ref{reference_measurements}, the details of the satellite receivers and the reference plasma density measurements are presented. In Section \ref{section_methodology}, we describe
our methodology for determining plasmaspheric electron densities.
In Section \ref{section_results} we demonstrate the applicability of our method on a selected subset of measurements; in Section \ref{sources_of_error} we briefly discuss the sources of error, and Section \ref{section_conclusion} summarizes our results.

\section{Observations and Datasets}

\label{section_observations}

\subsection{The Alpha Transmitters}

\label{section_transmitters}

Although the Alpha system has been in operation since the 1970's, to this day very little information has been published about it. We consulted \citet{jacobsen2006}, \citet{jaatinen2011} and \citet{balov2016radio} for basic information. In addition, we made VLF measurements at three AWDANet VLF receiver ground stations \citep{lichtenberger2008,lichtenberger2010}: at Tihany, Hungary; Tvarminne, Finland; and Karymshina in Kamchatka, Russia (all within 2000~km distance of an Alpha transmitter) to confirm the Alpha transmission sequences and to determine their exact timing.

\begin{figure}[!h]
        \includegraphics[width=17cm]{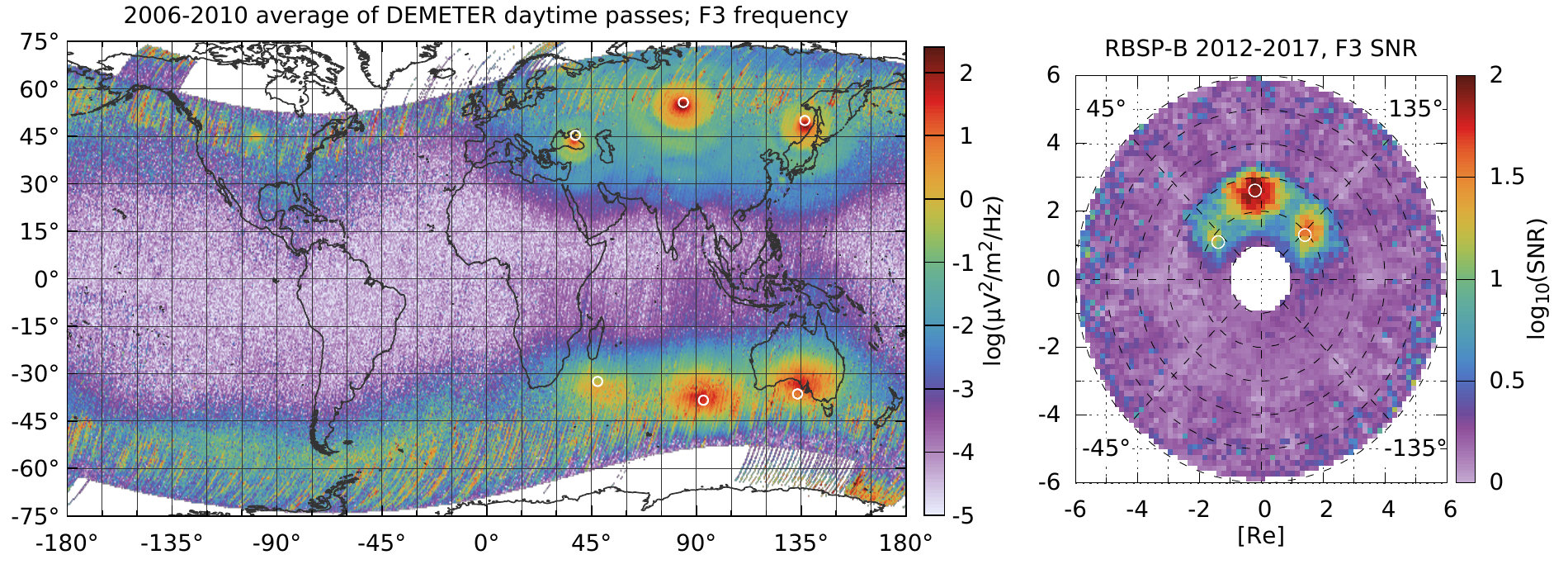}
        \caption{Left: Average electric field in the F3 alpha frequency above background levels. Long term average based on the ICE instrument of the DEMETER satellite in low Earth orbit, calculated from daytime passes of the Sun-synchronous satellite orbit. White circles are the locations of the three main Alpha transmitters (northern hemisphere, from Krasnodar, Novosibirsk and El'ban, west to east) and their magnetic conjugates (southern hemisphere). Right: Average signal to noise ratio of the F3 alpha frequency in the equatorial plane, based on HFR measurements of the EMFISIS instrument onboard the RBSP-B satellite. White circles denote the intersection of the plane and the magnetic field lines starting at the transmitters. The prime meridian is to the left.}
        \label{fig:demeter}
\end{figure}

The Alpha (RSDN-20) radionavigation system consists of 3-5 stations, each station transmitting a unique repeating sequence, a combination of short pulses chosen from 4 fixed frequencies (or an equal timespan of silence). The pulses are 400~ms long followed by 200~ms spacing. The sequences consist of 6 pulses, thus, the total length of a sequence is 3.6 seconds, with 1000 repeats per hour. This synchronization is disrupted at times of UTC leap seconds and re-established a couple of days later, according to our measurements. The main frequencies used are F1=11904~Hz, F2=12648~Hz and F3=14880~Hz. A fourth frequency, F4=12090~Hz is only used by the Revda station. The pulses contain no modulation or other information. The transmission sequence is shown in Table \ref{table:alphasequence}. As confirmed by our ground-based measurements, currently three transmitter stations are in regular operation (see map in Figure \ref{fig:demeter}), which are located in Krasnodar (45.403$^{\circ}$N 38.158$^{\circ}$E, L=1.79), Novosibirsk (55.758$^{\circ}$N, 84.446$^{\circ}$E, L=2.69) and El'ban (50.072$^{\circ}$N 136.609$^{\circ}$E, L=1.97) (the latter is also variously referred to in the literature as Khabarovsk or Komsomolsk-na-Amure (also Komsomolsk-on-Amur or Komsomolskamur), after the nearest main cities). A fourth station, located in Revda (68.037$^{\circ}$N 34.679$^{\circ}$E, L=5.56), has been active only for short periods of time. The literature lists a fifth station which we did not find any signs of being operational. More detailed descriptions of the system can be found in \citet{jacobsen2006} and \citet{jaatinen2011}.

\begin{figure}[!h]
        \includegraphics[width=9cm]{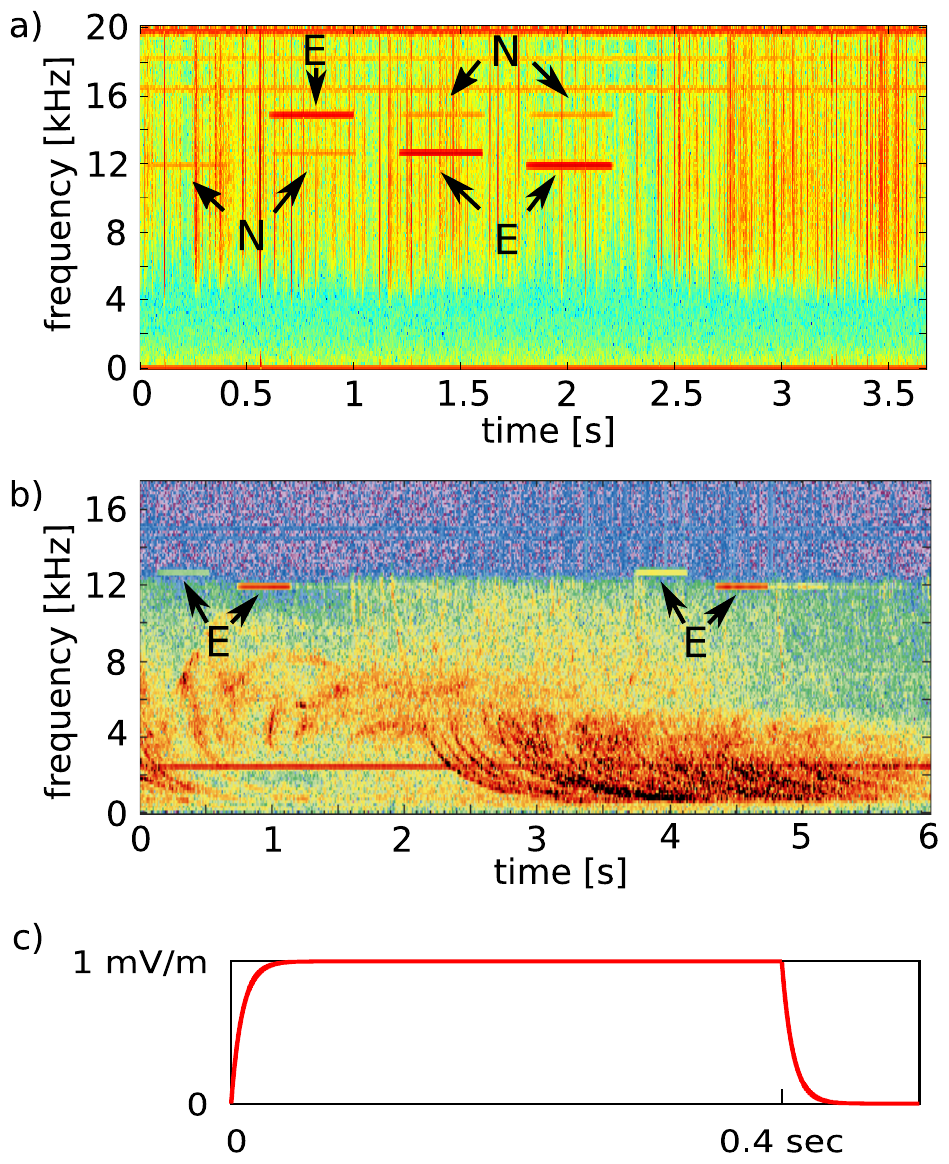}
        \caption{
Example of detected Alpha signals. (a) Spectrogram of a VLF sample recorded at the Karymshina ground station, Kamchatka. Signal sequences originating from El'ban (1500 km distance) and Novosibirsk (4500 km distance) are visible - c.f. table \ref{table:alphasequence}. 
As expected, signals from the closer station are stronger.
(b) Spectrogram of a VLF sample recorded by the RBSP-A satellite at a meridian near that of El'ban. The lower two frequencies of the transmission from El'ban are visible. Attenuation above 12~kHz is due to the analog filter of the instrument.
Signals from the other transmitters, located on different meridians, are absent.
(c) Signal envelope of an Alpha pulse, as determined by our ground-based measurements. The time constant of the exponential edges is 2~ms (exaggerated on the figure for better viewing).
}
        \label{fig:registratum}
\end{figure}

Each station emits a unique sequence of pulses and thus, when a VLF receiver clearly records a complete sequence, the transmitting station can be easily determined. 
In our test measurements with ground-based receivers, the relative power, the direction (in case of directional receivers), and the transmitter-to-receiver propagation delay of the pulse sequences corroborated their inferred sources.
The top panel of Figure \ref{fig:registratum} presents an example of a sequence recorded on the ground close to one of the transmitters (El'ban), showing a strong signal from the closer transmitter and a weaker signal from a more distant transmitter, while the middle panel is a satellite record made at the same meridian (El'ban), with the observable sequence consistent with the source being El'ban and not consistent with either of the other stations, c. f. Table \ref{table:alphasequence}.
If only a part of a sequence or only a single pulse is recorded, which is the case for the satellite measurements in this study, the spacecraft position and a reasonable modelling of propagation delay still allows us to ascertain the source transmitter. For the source to be misidentified, the reconstructed pulse timing would have to be off by at least half of the pulse slot length, or 300~ms. 

\begin{table}[hbt]
\caption{Transmission sequence of the three main Alpha stations (Krasnodar, Novosibirsk and El'ban) and the rarely utilized fourth station (Revda). The complete sequence is 3.6 seconds, divided into six 600~ms slots. The pulses themselves are 400~ms long followed by 200~ms gap.}

\begin{tabular}{ l c c c c c c }
\hline
 Pulse slot \# & \#1  & \#2 & \#3 & \#4 & \#5 & \#6 \\
\hline
Frequency    &  &  &  &  & - & - \\
  F3 = 14880 Hz  & Krasnodar & El'ban & Novosibirsk & Novosibirsk & - & - \\
                 &           &        & (Revda)     &             & & \\
\hline
  F2 = 12648 Hz & (Revda)   & Novosibirsk & El'ban & Krasnodar   & - & - \\
 \hline
 F4 = 12090 Hz & -          & (Revda)     & -      & -           & - & - \\
 \hline
  F1 = 11904 Hz & Novosibirsk & - & Krasnodar & El'ban           & (Revda) & - \\
\hline

\end{tabular}

\label{table:alphasequence}
\end{table}

For a given location in the magnetosphere, the local electron gyrofrequency can be calculated as $ f_{ce}=\frac{eB_0}{2\pi m} $, where $B_0$ is the local background magnetic field strength, $e$ and $m$ are the electron charge and mass, respectively.
For a ducted monochromatic signal to propagate along the entire length of a magnetic field-line, it is believed that its frequency should not exceed the half of the gyrofrequency at any point. Thus, for each L-shell and corresponding $B_{0min}$, there is a maximum frequency that can undergo interhemispherically ducted propagation - or conversely, for each frequency, there is an outer L limit within which such propagation occurs \citep{clilverd2008}. For instance, for F3=14.8 kHz, this outer limit is L=3.1, while for F1=11.9 kHz, it is L=3.3. Thus, we expect to observe ducted Alpha pulses below this L-value, and focus our investigations to measurements within this region. The actual geographic location of the three main transmitters happen to fall well inside this limit. It should be mentioned that even outside this L-limit, ducted propagation may occur on a section of a field line starting on the source hemisphere and ending somewhere short of the magnetic equator.

\subsection{The EMFISIS Instrument}
\label{emfisis}
The Van Allen Probes (earlier Radiation Belt Storm Probes, RBSP) satellite pair orbits Earth in a 10 degree inclination quasi-equatorial plane with 600 km perigee and 30000 km apogee.
EMFISIS is a wave instrument operating on these satellites. It records spectra and waveforms in 6 channels, from 3 magnetic search coils and 3 electric antennas. Its Waveform Receiver (WFR) can be operated in continuous burst mode to record 6 seconds (at a time) of continuous waveforms in all 6 channels, at 35 kHz sampling rate. Within the instrument, waveforms are passed through a 10 Hz - 12 kHz bandpass filter before digitization and recording \citep{kletzing2013}. Unfortunately, this filtering completely attenuates the F3 alpha frequency. However, F1 pulses, and due to the slow roll-off of the filter, occasionally F2 alpha pulses too, can be observed in the recordings. Thus, the 6 second long, 6-channel WFR continuous burst mode data are ideal for the accurate detection of one or more 400~ms Alpha pulses. Figure \ref{fig:registratum}b shows an   example of such a detection. The 6 channels also allow the determination
of wave characteristics, as we later show.

\subsection{Reference Measurements}
\label{reference_measurements}
Another part of the EMFISIS suite, the High Frequency Receiver (HFR) analyses the 10 kHz - 500 kHz range, and in its survey mode, it regularly records spectra of that range. These spectra are bin-averaged into 82 logarithmically spaced bins. These series of spectra are analyzed semi-automatically for the presence of plasma upper hybrid resonances to determine the local electron densities ("AURA" algorithm) and the results are regularly published as L4 data \citep{kurth2015}.
\shortciteA{zhelavskaya2016}
present a neural network trained on this same dataset which therefore yields very similar results without human oversight ("NURD" algorithm).
In cases where data from AURA is not yet available, which is the case in Figures \ref{fig:inverzio1}-\ref{fig:inverzio2}, NURD can be used instead.
Both of these datasets have an upper density limit of $n_e = 3000~ \mathrm{cm}^{-3}$.
We use these
official and published
datasets as a reference to validate our results.

The resolution of the densities obtained by AURA (and NURD) is 10\% ($\Delta n/n$). To improve on this, we processed 
the 4~ms long survey mode HFR wave measurements taken regularly at roughly 60~s time cadence and manually determined the upper hybrid frequency and the plasma frequency where possible. This method usually yields more accurate density values
($\pm2$ to 50 $\mathrm{cm}^{-3}$, depending on the signal clarity)
at a sparser sampling in time.
We
carried out this procedure for time periods presented in Figures \ref{fig:inverzio1}-\ref{fig:inverzio3} and
used this new dataset as additional reference
(shown on the figures).

\section{Methodology}
\label{section_methodology}
\subsection{Confirmation of ducted propagation} %\label{section_ducted}

In Section \ref{section_transmitters}, we argued that Alpha signals are expected to propagate in ducted mode, starting from or near the geographic location of the transmitters.
This is supported by measurements, as shown on Figure \ref{fig:demeter},
where we plotted the spatial distribution of long term average fields.
Left is a map of the long term average of the electric field at
one of the Alpha frequencies based on data from the
ICE instrument \citep{DEMETER_ICE} onboard the DEMETER satellite in low Earth Sun-synchronous orbit.
Excess power can be seen above the three main transmitters,
and importantly, also around their magnetic conjugates on the opposite hemisphere.
This is explained by signal propagation outside
the Earth-ionosphere waveguide,
ducted
along the magnetic field lines in the magnetosphere.
Such ducted signals
have also been directly observed on the ground in Australia near
one of the conjugate coordinates by
\shortciteA{tanaka87}
and
\shortciteA{cohen2009}
(see Figure 7 therein). On the right of Figure \ref{fig:demeter} we present a map of the average signal to noise ratio in the equatorial plane in Earth fixed polar coordinates (geographic longitude and radius), based on the EMFISIS HFR measurements on the RBSP-B
(one of the two Van Allen Probes) satellite.
This equatorial plane intersects the three magnetic field lines corresponding to the ground transmitters,
at roughly midway along the field lines.
Around the three intersection points on the plane, three
patches of stronger signals can be seen.

Due to different transmitter and instrument duty cycles, measurement
periods, etc. the signal strengths are not directly comparable between the two maps.
Nevertheless, their spatial distributions support ducted propagation.
The right panel of Figure \ref{fig:demeter} is also consistent with the L=3.1 upper limit of the ducted propagation corresponding to the relevant frequency, as discussed earlier.

\subsection{Determining Pulse Time of Arrival} \label{section_toa}

The fundamental input parameter of the inversion process
is the travel time of the signal from the transmitter on the ground
to the detectors on the spacecraft.
This requires that we 1) know the timing of the pulses emitted
by the transmitters on the ground and 2) measure the pulse time of arrival on the spacecraft. Since 1) is not
documented in the literature, we had to first determine
the precise timing of the emissions.
This was done by determining the pulse arrival times
at nearby ground-based VLF receivers
and calculating the timing of the emitted
signals by taking into account the propagation time in the Earth-ionosphere waveguide from the transmitter to the receiver.

As part of a hyperbolic navigation system, the Alpha pulses were designed for the detection and comparison of the phase of two pulses detected simultaneously.
It was never intended for the precise determination of the arrival
time of a single pulse. Nevertheless, it is still possible to determine the time of arrival with reasonable
precision, by comparing the pulse envelope with the envelope
of the transmitted signal, if the latter is known.
\citet{brown1977} discusses the same problem in the context of the similar Omega system.

To determine the timing of the emissions and the pulse envelope shape, we 
averaged successive pulses recorded at a fixed location on the ground.
We obtained the timing and the envelope to better than 1~ms precision (Figure \ref{fig:registratum}c).
The determined signal envelope was subsequently matched to
the signals detected on the RBSP satellites to determine the time of arrival.

When determining the time of arrival at the satellite,
the attainable accuracy varied for each signal.
For some pulses, it could be as good as 1~ms. However,
due to usually low signal-to-noise ratio and occasional deviations
from the established signal envelope shapes, accuracy was more typically between 1 and 10~ms. In some cases, no useful timing data could be obtained due to the signals being extremely faint or distorted. This may be a result of wave-particle or wave-wave interactions, attenuation, excitation, or different propagation (for example, multipath propagation may cause a stretching of the recorded signals, also observed by \citet{sonwalkar94}). We excluded such signals from our analysis.

\subsection{Wave Propagation Inversion} \label{section_inversion}

The total propagation time of a fractional-path whistler-mode wave from the source to the receiver at the satellite is $T = T_{wg} + T_i + T_m$, where $T_{wg}$ is the travel time in the Earth-ionosphere waveguide between the transmitter and the presumed exit point, $T_i$ is the travel time through the ionosphere and $T_m$ is the travel time in the plasmasphere medium.
We measure $T$ as discussed in Section \ref{section_toa}, and subtract $T_{wg}$ and $T_i$ to obtain $T_m$.

Subsequently, $T_m$, or the propagation in the plasmaspheric path is inverted using 
wave propagation model based on Appleton-Hartree dispersion relation
for longitudinal propagation, without approximation
(see e.g. \citet{helliwell1965}). 
The procedure is similar to the whistler inversion method for the monitoring of plasmaspheric electron densities described in \citet{lichtenberger2009}.
The  method to derive the plasma density from the whistlers, which we call a whistler inversion method, consists of three components: (1) a wave propagation model -- the cold plasma (Appleton-Hartree) dispersion relation for longitudinal propagation, (2) a magnetic field model (dipole or IGRF model), and (3) a plasma density distribution model along the propagation path (Ozhogin-model).
The travel time for the magnetospheric part for frequency $f$ is calculated as
(in a slightly simplified form, for full treatment, see \citet{lichtenberger2009})
\begin{equation}
 T_m(f)=\frac{1}{2c}\int_{path}{\frac{f_pf_H}{f^{1/2}(f_H-f)^{3/2}}ds},
\label{eq:1}
\end{equation}
where $f_p$ is the electron plasma frequency, $f_H$ is the local electron gyrofrequency and the integral is calculated over the propagation path. In the case of whistler inversion, this is the path along the field line from the top of the ionosphere at one hemisphere to the top of the ionosphere on the other hemisphere. The measured and calculated travel times are compared using a standard optimization procedure until we obtain the optimal parameters. The parameter to be optimized are the epoch of the causative sferic, the propagation L-value and the equatorial electron density,
There are two differences in the procedure applied to the Alpha signal compared to the one in whistler inversion: (1) the propagation L-value is known and (2) the integral
in Eq. \ref{eq:1} is calculated from the top of the ionosphere (taken as 2000~km)
to the magnetic latitude of the satellite. Because of the latter condition, this procedure is called fractional-hop inversion.
In this analysis we used the IGRF-12 geomagnetic model \citep{IGRF12}
and the \citet{ozhogin2012} density profile for this purpose.
The inversion procedure yields an equatorial electron density,
which can also be easily converted into local electron density at the satellite location for direct comparison with the in-situ reference measurements.

The field-aligned density profile model used here was developed using 700 active sounding experiments by the RPI instrument on the IMAGE satellite made between 2000 and 2005. The remote sounding provides practically instantaneous high precision field-aligned density distributions, which were statistically analyzed to develop a general empirical model formula with few parameters. It was shown that while significant variations exist, it is currently the best model of the latitude dependence of the plasmaspheric density along the field lines \citep{ozhogin2012}.

We estimated $T_{wg}$ as the geodesic distance between the spacecraft footprint and the transmitter divided by the speed of light.
Actual propagation in the waveguide is slightly slower but this
causes a negligible (1 to 10~$\mu$s) difference in time \citep{helliwell1965}.
For the propagation time across the ionosphere, $T_i$,
we relied on the formula of \cite{park1972}, which depends on the critical frequency of the F2 layer, $f_0\textrm{F}_2$.
To get an approximation of the ionosphere at every satellite footprint coordinate,
we calculated $f_0\textrm{F}_2$ values using the IRI-2016 (International Reference Ionosphere, \citet{IRI2016}) model, and rescaled those values by actual measurements at a couple of locations.
These ionospheric measurements were taken from nearby ionosonde stations, which were the following: Beijing, I-Cheon, Khabarovsk, Kokubunji, Magadan, Manzhouli, Mohe, Petropavlovsk/Paratunka, Wakkanai, Yakutsk (shown as yellow squares on the inset maps on Figures \ref{fig:inverzio1}-\ref{fig:inverzio3}).

\section{Results} \label{section_results}

Here we present an analysis of a subset of the measurements of our 2016 EMFISIS
campaign targeting the Alpha transmitters. We selected three 30-minute periods
(starting at 2016-02-15 05:15 UT, 2016-03-18 17:00 UT and 2016-03-24 18:00 UT) for a particularly consistent presence of Alpha pulses.
The first and last sets were obtained by turning on the continuous burst mode measurement
for 1 minute every 3 minutes. In the case of the second set, 6-second long continuous burst measurements
were triggered intermittently by power in the VLF band (possibly due to the presence of strong whistlers), and thus this sequence is more irregular.

These three sets of detections were chosen for their quality, that is, the presence of a large number of Alpha signals at good signal to noise ratio.
Nevertheless, during these periods, we also observed a couple of non-detections, or pulses that were missing, or possibly swamped by the noise.
In addition, we also observed a small number of pulses with hardly discernible or distorted envelope shapes which prevented the determination of their timing; some overlapping pulses; and pulses with indeterminable propagation directions.
These may possibly represent a lack of ducting structures, pulses attenuated by the medium or pulses undergone wave-particle interactions, multiple path pulses and obliquely propagating pulses, respectively.
Such signals were not part of our analysis and are not shown on our figures, for sake of clarity.
The only exception is the sequence of obliquely propagating pulses in Figure \ref{fig:inverzio3}.

\begin{figure}[p]

    \includegraphics[width=14cm]{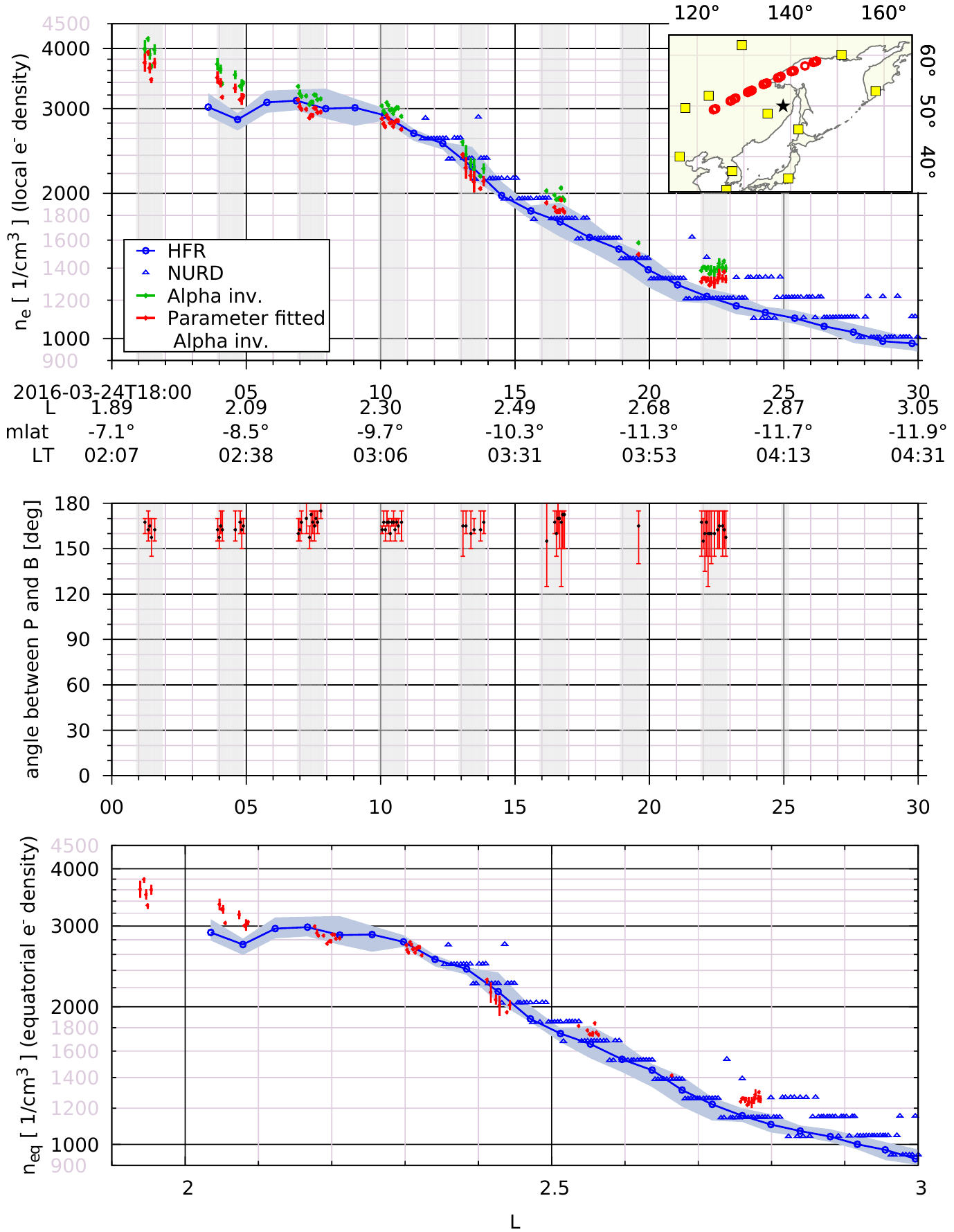}
    \caption{Top: Comparison of local electron densities obtained with our VLF signal inversion method (green dots - using \citet{ozhogin2012} model with default parameters, red dots - with fitted parameters; error bars shown stem from uncertainty of pulse arrival time determination) to densities from upper hybrid frequencies as a reference (blue triangles - NURD, blue circles - HFR analysis, blue shaded area - uncertainty).
Shaded gray areas are the periods when EMFISIS WFR
continuous burst
mode measurements were turned on.
Inset: map of satellite footprints (red circles), source transmitter (star shape) and ionosonde stations (squares).
Middle: angle of Poynting vector of each signal with respect to the field lines.
Bottom: comparison of inversion results and reference measurement, both converted to equatorial electron densities, against L-value. Only the parameter fitted results are shown.
}
     \label{fig:inverzio1}
\end{figure}

\begin{figure}[p]

    \includegraphics[width=14cm]{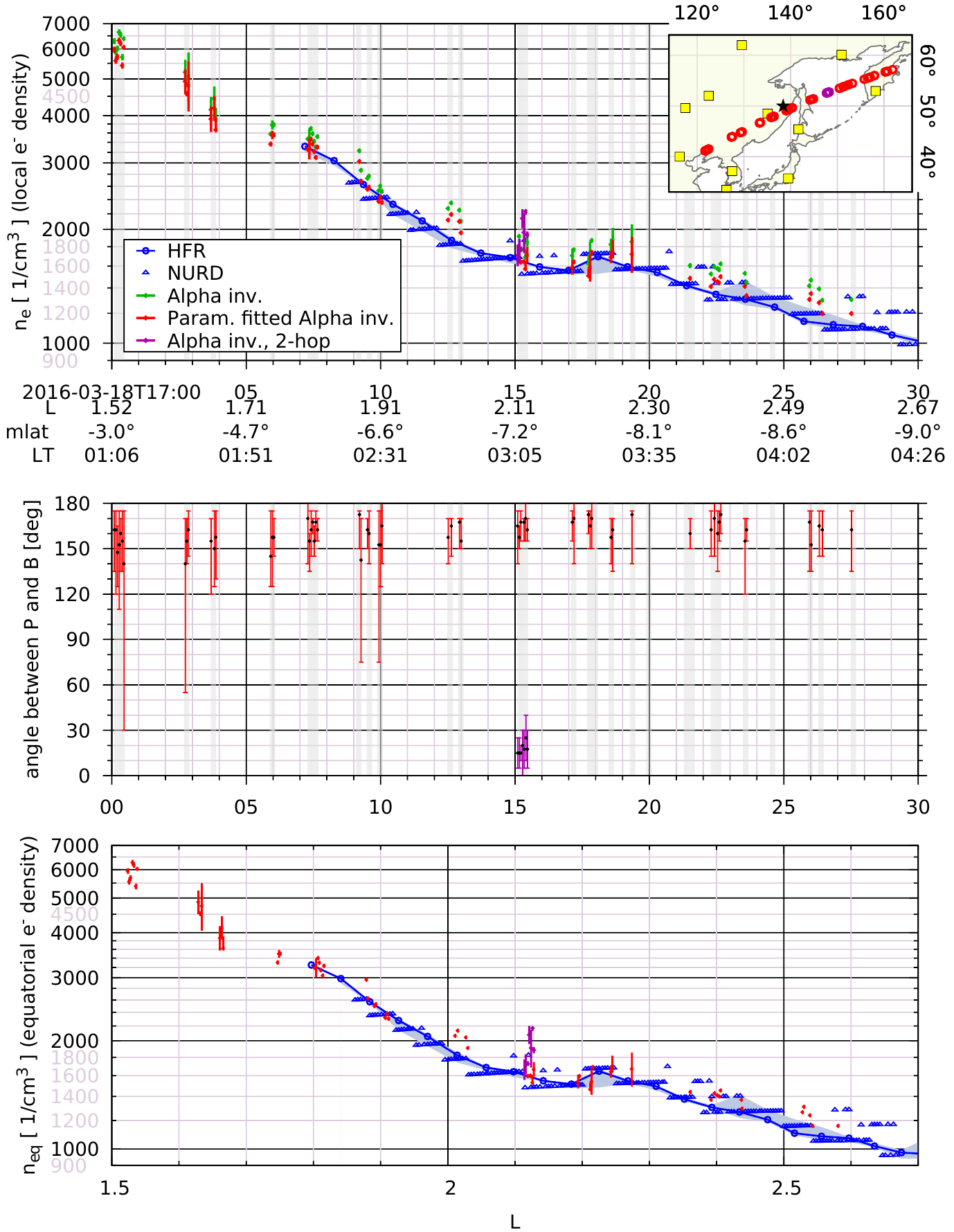}
    \caption{Panels are same as in Figure \ref{fig:inverzio1}.
Some Poynting vectors (middle panel, purple bars) show parallel propagation in the opposite direction - that is, toward the transmitter. This is a result of signal reflection from the southern hemisphere ionosphere, similar to 2-hop whistlers. Inversion results of these signals, assuming such 2-hop propagation (and using the fitted \citet{ozhogin2012} parameters), are shown in purple.}
    \label{fig:inverzio2}
\end{figure}

\begin{figure}[p]

    \includegraphics[width=14cm]{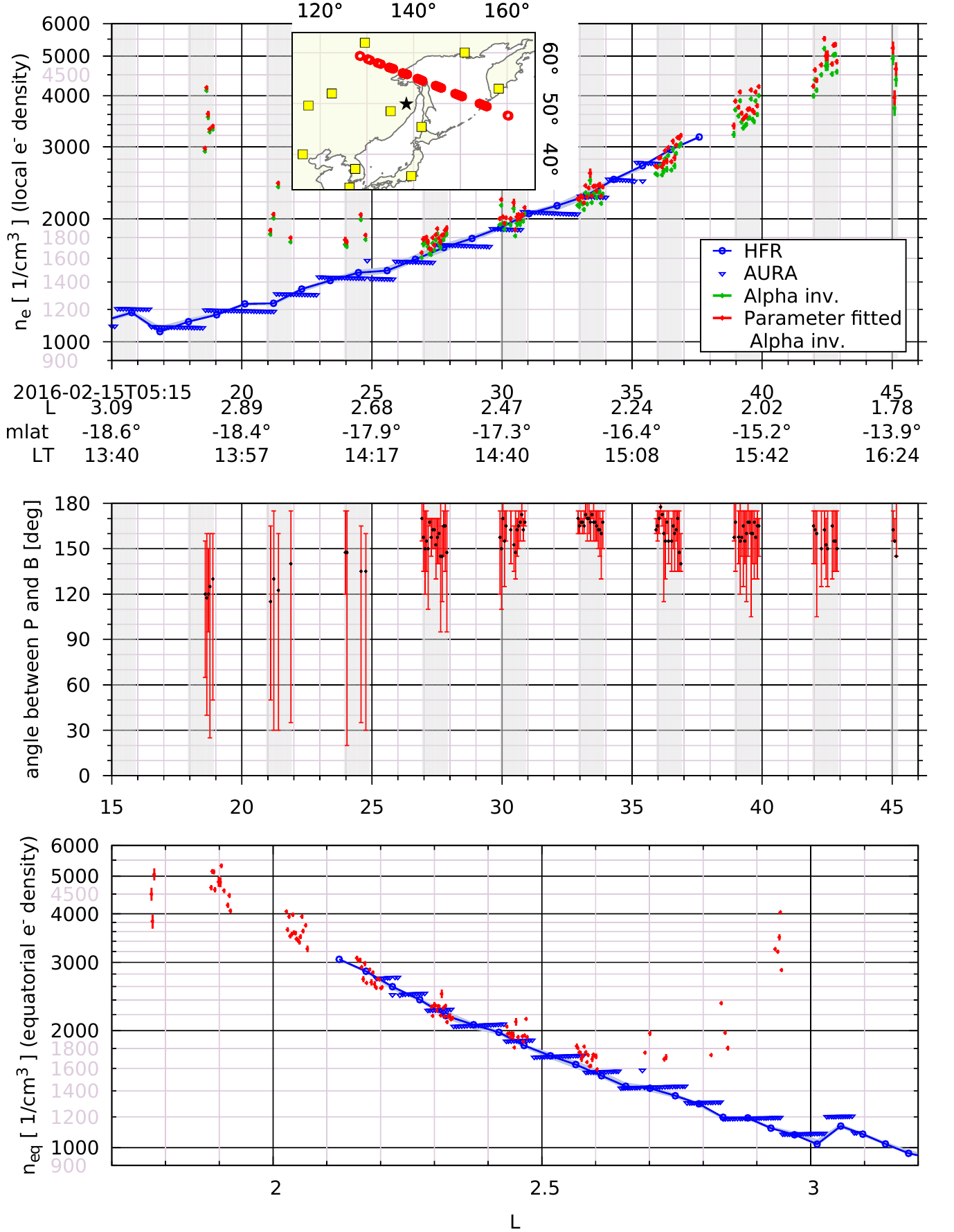}

\caption{Panels are same as in Figure \ref{fig:inverzio1}, but reference densities are from AURA (blue triangles).
Here the first three groups significantly deviate from parallel propagation, which explains the disagreement between inversion results and reference measurement (outliers on top \& bottom panel). Note that during this observation the satellite was descending in altitude (as opposed to those on Figures \ref{fig:inverzio1}-\ref{fig:inverzio2}) and thus the density was increasing with time, as can be seen on the top panel.}
    \label{fig:inverzio3}
\end{figure}

To process the data, first we identified any possible Alpha pulses at the F1 frequency,
measured the timing of each pulse, and carried out the inversion process
assuming field-aligned propagation
as discussed in Section \ref{section_inversion}.
With the aid of the underlying field-aligned density profile model \citep{ozhogin2012},
we can obtain a density value at any point along the field line.
On the top panels of Figures \ref{fig:inverzio1}-\ref{fig:inverzio3},
we compare the densities obtained at the location of the satellite to
in-situ measurements of different nature (based on the detection of upper hybrid
resonances in the plasma) as a reference.
Altogether 233 pulses were processed.

While \citet{ozhogin2012} give an average model of the field-aligned density profile,
their individual measurements show considerable variation, which can be accommodated
for by changing the parameters ($\alpha=1.01$, $\beta=0.75$) of their analytic formula:
\begin{equation}
N(L,\lambda) = N_{eq}(L) \cos^{-\beta} \left( \frac{\pi}{2} \frac{\alpha\lambda}{\lambda_{\textrm{INV}} } \right)
.\end{equation}
We searched pairs of $\alpha$ and $\beta$ parameters
that lead to a better match between our inversion and the reference measurements.
Indeed, we ended up with three parameter pairs that slightly
deviate from the values in the average model,
and give a better match to our three datasets
(2016-02-15: $\alpha=1.00 \pm 0.05$, $\beta=0.25 \pm 0.05$;
2016-03-18: $\alpha=1.10 \pm 0.05$, $\beta=0.75 \pm 0.05$;
2016-03-24: $\alpha=1.05 \pm 0.05$, $\beta=0.90 \pm 0.05$).
These values are within the ranges observed by \citet{ozhogin2012}.
Inversion results with both the average model and the parameter fitted models
are presented in the top panel of Figures \ref{fig:inverzio1}-\ref{fig:inverzio3}.
The scatter in the density values is due to different
pulse arrival times, which may be either real or due to measurement error, as discussed later.

As a second step, using the full, 6-component electromagnetic wave measurement
of the EMFISIS instrument, we calculated the spectral density of the modulus of the Poynting
vector using the real parts of the cross-power spectra between the
6-channel components \citep{santolik2010}.
This is to confirm that the Poynting vectors are parallel to the field line and thus the propagation was ducted.
The middle panels of Figures \ref{fig:inverzio1}-\ref{fig:inverzio3}
show the obtained angles between the field line and the Poynting vectors,
with 180$^{\circ}$ corresponding to parallel propagation away from the transmitter.
Error bars represent the full width at half maximum (FWHM)
of a gaussian fitted to a histogram of the pixel values
corresponding to the recorded signal on the Poynting spectra.
Error bars on Figure \ref{fig:inverzio3} are somewhat larger than on Figures \ref{fig:inverzio1}-\ref{fig:inverzio2} due to
larger background noise on the Poynting spectra.
We did not correct for the effect of any background noise
on the determination of Poynting vectors, which may shift the
results toward 90$^{\circ}$. Even without further correction, the obtained
values are sufficiently good to decide between ducted and unducted propagation.

Indeed, the propagation direction of most of the signals was close to 180$^{\circ}$, confirming ducted propagation.
An exception is the signals before the t=25 minute mark on the top panel of 
Figure \ref{fig:inverzio3}, which exhibit oblique propagation.
(Such oblique propagation of VLF waves is not uncommon in the magnetosphere, see for example \citet{zhang2018}.)
Accordingly, as expected, inverting these signals with our procedure, which assumed ducted propagation, lead to strong disagreement with the reference values.
This underlines how in the rest of the cases the obtained propagation direction and
the agreement between the inversion results and the reference support each other.

In the middle panel of Figure \ref{fig:inverzio2}, a set of pulses close to 0$^{\circ}$, 
can be observed, signifying propagation
parallel to the field line but in the opposite direction, \textit{toward} the transmitter.
These are signals that traveled all the way to
their conjugate points, were reflected from the ionosphere there, and
returned along the same path, where the satellite recorded them.
The same process is responsible for the well known whistler echoes identified as 2-hop whistlers \citep{helliwell1965}.
Luckily, our inversion method can be carried out for these signals too,
by assuming ducted propagation for both the first and the return path.
The results of the inversion of these return signals are shown in the top/bottom panels of Figure \ref{fig:inverzio2},
and are in relatively good agreement with the reference measurement.
This fact again supports the theory of ducted propagation.

Finally, the bottom panels of Figures \ref{fig:inverzio1}-\ref{fig:inverzio3}
serve as a final comparison of our results to the reference measurement. 
Here we converted the in-situ reference measurement to equatorial electron
densities and the densities from the inversion are also evaluated at the equator and plotted against the L-value.
For clarity, only the inversion results with the fitted parameter \citet{ozhogin2012} models are shown.

\section{Sources of Error}
\label{sources_of_error}

The two main sources of errors in the determined density values are
the uncertainty of the measured travel time of the VLF signals
and the assumptions involved in the inversion process.
The former involves uncertainties in both emission and arrival time.
Since we have determined the signal emission timing to better than 1~ms precision,
the main source of uncertainty here is the uncertainty of the arrival time on the spacecraft.
This depends on the clarity and sharpness of the signals and therefore the signal to noise ratio.
It is usually under 10~ms, and in some cases can be as good as 1~ms.
These uncertainties are shown as error bars on the inversion results
in our figures (top and bottom panels of Figures \ref{fig:inverzio1}-\ref{fig:inverzio3}).

Unfortunately, 
the satellites have no onboard
high precision time signal source such as GPS receivers, and thus
the timestamps in the EMFISIS waveform
data will have their own uncertainty
with regards to actual UTC.
The original EMFISIS design requirement called for a post-processing accuracy of 50~ms
(see e.g. \citet{kirby2013}; note that the latter document contains a couple of erroneous values regarding the
components of timing, as confirmed by the authors. Correct values can be found in \citet{kirby2012}.).
Despite this broad requirement, our correspondence with the engineers involved suggests
that the error of absolute times in the EMFISIS data is probably much better, on the order of
5~ms.
This uncertainty, though, still precludes us from determining with certainty whether the observed scatter
in the arrival times between consecutive signals (which can be seen
as an occasional scatter of
consecutive density values from the Alpha inversion
in Figures \ref{fig:inverzio1}-\ref{fig:inverzio3}),
is due to measurement error or due to actual variation in the propagation of consecutive signals.

The uncertainty in the ionospheric density can affect the inversion results.
However, the typical ionospheric contribution to the travel time of the signals (the ratio of $T_i$ to $T_m$) is low. In the case of our three sets of measurements it is as follows: 3-20\% (2016-02-15), 5-25\% (2016-03-18) and 3-10\% (2016-03-24) with larger values corresponding to lower L-shells. This in turn means that even 50\% error in the $f_0\textrm{F}_2$ critical frequency leads to 3-10\%, 4-20\% and 3-7\% error in the obtained densities, respectively.

Note that the error bars of the top/bottom and middle panels on Figures \ref{fig:inverzio1}-\ref{fig:inverzio3} have a different origin.
The error bars of the plasmaspheric densities
stem from the uncertainty in the pulse time of arrival of the observed Alpha signals,
i. e. the sharpness of the signal edges, which may depend on various effects of noise during the entire propagation,
and are also affected by the L-value and the value of the density itself by way of the inversion procedure.
On the other hand, the error bars on the middle panels of Figures \ref{fig:inverzio1}-\ref{fig:inverzio3} depend on the local noise conditions 
(i. e. background noise and possible other VLF noise sources such as whistlers, co-temporal with the signal envelopes)
in the Poynting spectra calculated from all 6 channels.
Thus, the two kind of errors are of different source and nature and do not necessarily vary in a correlated way.

\section{Conclusion}
\label{section_conclusion}

We have detected VLF pulses on the Van Allen Probes satellites originating from the Russian RSDN/Alpha transmitters. We have shown that several sequences of these signals have undergone ducted propagation, based on both their timing and their Poynting vectors. We applied propagation inversion to the measured signals, yielding plasmaspheric electron densities along the respective geomagnetic field lines. These values, when taken at the location of the satellites, were shown to be in good agreement with the local densities determined by an independent method that relies on the local upper hybrid resonance in the plasma.
Furthermore, in some cases, we also detected echo signals, reflected from the opposite hemisphere and returning along the same field line all the way to the satellite. The inversion along their full path, including the return path, again lead to a good agreement with the rest of the measurements.
Thus, the results of our procedure supports the validity of the propagation inversion, and the underlying assumptions: ducted propagation along the field line, and the
underlying field-aligned density profile model \citep{ozhogin2012}.
We have shown that a slight tuning of the density profile model parameters, within its
suggested ranges, improves the agreement between our inversion results and
the independent reference measurement.

We have also observed several non-detections. This is consistent with the theory of whistler propagation guided along ducts following geomagnetic field lines, when such ducts are present, and a lack of such signals at other times.

In principle, the detection and inversion of ducted VLF pulses outlined above can be viewed as an alternative method of density measurement in the plasmasphere, if the field- aligned density profile model is accepted as valid. Opposed to local measurements, this procedure is sensitive to the density along the complete propagation path.
Furthermore, this method may in principle extend the measurement ranges in density to values not covered by other instruments. For example, the density values determined from upper hybrid resonances on the Van Allen Probes satellites are limited to a maximum of about 3000~$\mathrm{cm}^{-3}$ due to the upper frequency limit of the wave measurements, which can be clearly exceeded by
our inversion results (Figures \ref{fig:inverzio1}-\ref{fig:inverzio3}).
Our method relies only on the precise time determination of the signals. In case of the Van Allen Probes satellites, it is in practice limited by the uncertainty of the wave measurement timestamps. Future satellites with more accurate time measurement may show the limits of our method with regard to precision and range. Thus, it may be worth considering
the opportunities offered by more accurate time measurement in the design of new wave experiments.

It is worth noting that while transionospheric propagation was considered a correction factor in our inversion, a similar method can be applied in principle to satellites in lower orbit, to probe the ionosphere itself.

\acknowledgments
The authors are grateful for the assistance
of Craig Kletzing and the EMFISIS team,
Jerry Needell and Stan Cooper for useful correspondence, Juha Jaatinen for information on the Alpha network and Veronika Barta for help in verifying ionograms.

We are also grateful to the Guest Investigator program issued by CNES for the DEMETER mission for supplying raw data.
The authors acknowledge the following data sources of ionospheric measurements:
Global Ionospheric Radio Observatory (GIRO);
SPIDR (Space Physics Interactive Data Resource) database;
WDC for Ionosphere and Space Weather, Tokyo, National Institute of Information and Communications Technology;
Institute of Cosmophysical Research and Radio Wave Propagation, Far Eastern Branch of the Russian Academy of Sciences;
Institute of Solar-Terrestrial Physics of the Siberian Branch of the Russian Academy of Sciences.

The research leading to these results received funding from the European Space Agency and the National Research, Development and Innovation Office of Hungary under grant agreements 4000115369, NN116408 and NN116446.
Parts of this work was supported by JHU/APL contract no. 921647 under NASA Prime contract no. NAS5-01072 and JHU/APL contract no. 131802  under NASA prime contract no. NNN06AA01C.

\end{document}